\documentclass[aps,preprint,superscriptaddress]{revtex4}

\usepackage{amsmath}
\usepackage[dvips]{graphics}
\usepackage[dvips]{graphicx}
\usepackage{psfrag}
\usepackage[mathcal]{eucal}

\begin{document}
% You should use BibTeX and apsrev.bst for references
\bibliographystyle{apsrev}

\preprint{COLO-HEP-457}
\preprint{February, 2001}

\title{Implications of the Renormalization Group Equations
in Three-Neutrino Models with Two-fold Degeneracy} 
% Optional argument for running titles on pages %\title[]{}



\author{Mu-Chun Chen}
\email[]{mu-chun.chen@colorado.edu}
\author{K.T. Mahanthappa}
\email[]{ktm@verb.colorado.edu}
%\thanks{}
\affiliation{Department of Physics, University of Colorado, Boulder,
CO80309-0390, U.S.A.}


%\date{\today}

\begin{abstract}
We obtain a {\it complete set} of one-loop RGE's for a set of combinations of
neutrino parameters for the case of two-fold degenerate hierarchical
three-neutrino models. The requirement of consistency of exact solutions to
these RGE's with the two-fold degeneracy yields 
conditions which have previously been obtained perturbatively/numerically.
These conditions, in the limit $|U_{e\nu_{3}}|=0$, are shown to lead to a
strong cancellation in the matrix element of neutrinoless double beta decay.
\end{abstract}  

%\maketitle must follow title, authors, abstract and \pacs
\maketitle


%\section{}
%\label{}
%\subsection{}
%\subsubsection{}

It has been a puzzle that the mixing in the leptonic sector is so large while
the mixing in the quark sector is so small. Many attemps have been made to
explain this fact. One possible scenario is to utilize the flavour symmetry
combined with GUT symmetry at some high energy scale $\Lambda$
\cite{Albright:2000sz}. %%,Chen:2000fp,Blazek:1999ue,Berezhiani:2000cg}%% 
Viable symmetries are those giving rise to large
mixing in the lepton mass matrices. Most models of this kind suffer from
fine-tunning and the difficulty of constructing a viable superpotential in the
flavour symmetry sector that gives rise to the required vacua. An alternative
to this scenario is the idea of infrared fixed point (IRFP) 
\cite{Pendleton:1981as}. %%,Hill:1981sq,Lanzagorta:1995gp}.%% 
Contrary to the idea of
flavour combined with GUT symmetry, in the IRFP scenario, the low energy
physics  is governed by the low energy dynamics, namely, the renormalization
group equations below the scale $\Lambda$. Physics above the scale $\Lambda$
plays no role in the predictions at low energies. Therefore, if there exists
any IRFP which leads to viable phenomenology, one does not have to deal with
the fine-tunning problem and the difficulty of finding the correct vacua. The
focus of our attention in this note is the hierarchical three neutrino models
with two-fold denegeracy, and the implications of the exact solutions to
one-loop renormalization equations (RGE's) to these models.

In the flavour basis where the charged leptons are diagonal, the neutrino
flavour eigenstates and mass eigenstates are related by  $|\nu_{\alpha}> = 
U_{\alpha i} |\nu_{i}>$, where $\alpha$ and $i$ are the flavour and mass
eigenstate indices respectively. The mass matrix $m_{\nu}$ can be diagonalized
as follows 
\begin{equation}
U^{T}m_{\nu}U = diag(m_{1},m_{2},m_{3})
\end{equation}
We adopt the usual parametrization for the leptonic mixing matrix $U$
\begin{equation}
U=diag(e^{i\delta_{e}},e^{i\delta_{\mu}},e^{i\delta_{\tau}}) \cdot V \cdot 
diag(e^{-i\phi/2},e^{-i\phi^{'}/2},1)
\end{equation}
\begin{equation}
V=\left(
\begin{array}{ccc}
c_{2}c_{3} & c_{2}s_{3} & s_{2}e^{-i\delta}\\
-c_{1}s_{3}-s_{1}s_{2}c_{3}e^{i\delta} &
c_{1}c_{3}-s_{1}s_{2}c_{3}e^{i\delta} & s_{1}c_{2}\\
s_{1}s_{3}-c_{1}s_{2}c_{3}e^{i\delta} &
-s_{1}c_{3}-c_{1}s_{2}s_{3}e^{i\delta} & c_{1}c_{2}
\end{array}
\right)
\end{equation}
where  $c_{i} \equiv \cos \theta_{i}$ and $s_{i} \equiv
\sin \theta_{i}$ and $0 \leq \theta_{i} \leq \pi/2$. The $\delta_{e,\mu,\tau}$
are three unphysical phases which can be absorbed by phase redefinition of the
neutrino flavour eigenstates. There are three physical phases: $\delta$ is the
universal phase (analog of the phase in the CKM matrix), and $\phi$ and
$\phi^{'}$ are the Majorana phases. By properly choosing the phases $\phi$ and
$\phi^{'}$ all three mass eigenvalues $m_{i}$ can be made positive.
We therefore assume, without loss of generality, that $(m_{1}, m_{2}, m_{3})$
are positive. If any of these three phases is not zero or not $\pi$, 
CP violation in the lepton sector is implied. Note that in the limit
$\theta_{2} = 0$, $\theta_{1}$ is identified as the atmospheric mixing angle,
$\theta_{atm}$, and $\theta_{3}$ is identified as the solar mixing angle,
$\theta_{\odot}$. In general, the mixing matrix elements are related to the
physical observables,  the atmospheric and solar mixing angles, by
$\sin^{2}2\theta_{atm} \equiv 4|U_{\mu 3}|^{2}(1-|U_{\mu3}|^{2})$, and
$\sin^{2}2\theta_{\odot} \equiv 4|U_{e2}|^{2}(1-|U_{e2}|^{2})$.
Recent results indicate~\cite{Gonzalez-Garcia:2000sq} that  
for atmospheric neutrino oscillations, 
$\Delta m_{atm}^{2} = 3.1 \times 10^{-3} eV^{2}$, 
$\sin^{2}2\theta_{atm}= 0.972$ \cite{Fukuda:1998mi}; 
for solar neutrino anomaly problem, there exists four solutions: 
(i)VO: $\Delta m_{\odot}^{2} =8.0 \times 10^{-11} eV^{2}$, 
$\sin^{2}2\theta_{\odot}=0.75$, 
(ii)LOW: $\Delta m_{\odot}^{2} = 7.9 \times 10^{-8} eV^{2}$, 
$\sin^{2}2\theta_{\odot}=0.96$, 
(iii)LAMSW: $\Delta m_{\odot}^{2} = 1.8 \times 10^{-5} eV^{2}$,
$\sin^{2}2\theta_{\odot}=0.76$,  
(iv)SAMSW: $\Delta m_{\odot}^{2} = 5.4 \times 10^{-6} eV^{2}$, 
$\sin^{2}2\theta_{\odot}=6.0 \times 10^{-3}$
\cite{solar:2000}
; and the matrix element 
$|U_{e3}|=\sin \theta_{2}$
is constrained by the CHOOZ experiment to be $|U_{e3}| < 0.16$
\cite{Apollonio:1999ae}.

The observed relation 
$\Delta m_{atm}^{2} \equiv |m_{3}^{2}-m_{2}^{2}| \gg \Delta m_{\odot}^{2} 
\equiv |m_{2}^{2}-m_{1}^{2}|$ in the two-fold degenerate, hierarchical model
implies $m_{3} \gg m_{2} \simeq m_{1}$, and this in turn implies 
$\nabla_{21} \gg \nabla_{32} \simeq \nabla_{31} \simeq 1$ with
$\nabla_{ij} \equiv (m_{i}+m_{j})(m_{i}-m_{j})^{-1}$.

We assume that the neutrino masses are generated by a dimension-5
effiective Majorana mass operator in the MSSM
\begin{equation}
\label{lagrangian}
\mathcal{L} \supset - k_{ij} (H_{u} L_{i}) (H_{u} L_{j}) + h.c.
\end{equation}
The neutrino mass matrix $(m_{\nu})_{ij}$ is related to 
$k_{ij}$ by $m_{\nu} = k_{ij} v^{2} / 2$, where $v^{2} \equiv
v_{u}^{2} + v_{d}^2 = (246 eV)^{2}$ is the squared vacuum expectation value
of the SM Higgs. The effective dimension-5 operator is generated by some
mechanism at the high energy scale $\Lambda$. The seesaw mechanism is the most
common way to generate this operator. Since we are only interested in physics
below the scale $\Lambda$, we will start with the effective 
Lagrangian Eq.~(\ref{lagrangian}) without specifing the origin of this
effective operator. 

The general one-loop RGE of the effective left-handed Majorana neutrino  
mass operator is
given by \cite{Chankowski:1993tx} %%,Babu:1993qv}%%
\begin{equation}
\label{rgem}
\frac{d m_{\nu}}{dt}=-\{\kappa_{u}m_{\nu}+m_{\nu}P+P^{T}m_{\nu}\}
\end{equation}
where $t \equiv \ln \mu$.
In the MSSM, $P$ and $\kappa_{u}$ are given by,
\begin{eqnarray}
P & = & -\frac{1}{32\pi^{2}} \frac{Y_{e}^{\dagger}Y_{e}}{\cos^{2} \beta} 
\simeq -\frac{1}{32\pi^{2}} \frac{h_{\tau}^{2}}{\cos^{2}\beta} diag(0,0,1)\\ 
\kappa_{u} & = &
\frac{1}{16\pi^{2}}[\frac{6}{5}g_{1}^{2} + 6g_{2}^{2} 
- 6 \frac{Tr(Y_{u}^{\dagger}Y_{u})}{\sin^{2}\beta}] \nonumber\\
& \simeq & 
\frac{1}{16\pi^{2}}[\frac{6}{5}g_{1}^{2} + 6g_{2}^{2} 
- 6 \frac{h_{t}^{2}}{\sin^{2}\beta}] 
\end{eqnarray}
where $g_{1}^{2}=\frac{5}{3}g_{Y}^{2}$ is the $U(1)$ gauge coupling
constant, $Y_{u}$ and $Y_{e}$ are the $3 \times 3$ Yukawa coupling matrices for
the up-quarks and charged leptons respectively, and $h_{t}$ and $h_{\tau}$ are
the SM $t$- and $\tau$-Yukawa couplings.  Since $\kappa_{u}$ gives rise to an
overall rescaling of the mass matrix, it has no effects on the running of
the mixing matrix $U$. Eq.~(\ref{rgem}) can be solved analytically by
integrating out its right-hand side  \cite{Ellis:1999my}. Note that at
one-loop level, since the evolutions of the gauge coupling constants
$g_{1,2}(t)$ and of the diagonal Yukawa couplings $h_{t,\tau}(t)$ are known, it
is indeed possible to carry out the integrations on the right-hand side without
making any further assumptions. However, the diagonalization procedure of the
resulting $3 \times 3$ complex symmetric matrix, $m_{\nu}(t)$, is very
complicated. It is thus hard to infer analytically the behaviours of the
physical observables, the mixing angles and phases. An alternative to
this``run-and-diagonalize'' procedure is the ``diagonalize-and-run''
procedure.  It is convenient to work with the RGE's of mass eigenvalues and the
diagonalization matrix, given by \cite{Casas:1999tg} 
\begin{equation}
\label{rge:m} \frac{d m_{i}}{dt}=-2m_{i}\hat{P_{ii}}-m_{i}Re\{\kappa_{u}\}
\end{equation} \begin{equation} \label{rge:u} \frac{dU}{dt}=UT 
\end{equation}
where
\begin{equation}
T_{ii} \equiv i\hat{Q_{ii}}
\end{equation}
\begin{eqnarray}
T_{ij} & \equiv &
( \frac{1}{ m_{i}^{2} - m_{j}^{2} } ) \{ ( m_{i}^{2} + m_{j}^{2} )
\hat{P_{ij}} + 2m_{i}m_{j} \hat{P_{ij}^{\ast}} \} + i \hat{Q_{ij}} \nonumber\\
 & = & \nabla_{ij} Re\{\hat{P_{ij}}\} + i \nabla_{ij}^{-1} Im\{\hat{P_{ij}}\} 
+ i\hat{Q_{ij}}
\end{eqnarray}
Here $\hat{P}$ and $\hat{Q}$ are defined as
\begin{equation}
\hat{P} \equiv \frac{1}{2} U^{\dagger} (P+P^{\dagger})U, \quad
\hat{Q} \equiv \frac{-i}{2} U^{\dagger} (P-P^{\dagger})U
\end{equation}
Eq.~(\ref{rgem}), (\ref{rge:m}), and (\ref{rge:u}) have been studied before 
\cite{Casas:1999ac,Casas:1999tg,%
Chankowski:1999xc,Haba:1999xz,Balaji:2000gd,Balaji:2000ma},
but the analyses have been done either numerically or perturbatively.
(Exact solutions to the RGE's in a two flavour case have been investigated
recently in \cite{Balaji:2000ma}). 
Due to the large interfamily hierarchy in the charged lepton sector, we keep
only the $\tau$-Yukawa coupling. We will further assume that $h_{\tau}$
does not evolve throughout the entire range of the RG running. This is a valid
assumption for the hierarchical case with two-fold degeneracy, as
$\nabla_{21}$ is very large. Under these assumptions, the above quantities are
given in terms of the masses and the diagonalization matrix elements as
\begin{equation}
\hat{P_{ij}}=-\frac{h_{\tau}^{2}}{32\pi^{2}}U_{3i}^{\ast}U_{3j}, \quad
\hat{Q}=0
\end{equation}

\begin{figure}
\psfrag{y}[][]{\small $\nabla_{21}(t)$}
\includegraphics[scale=0.75]{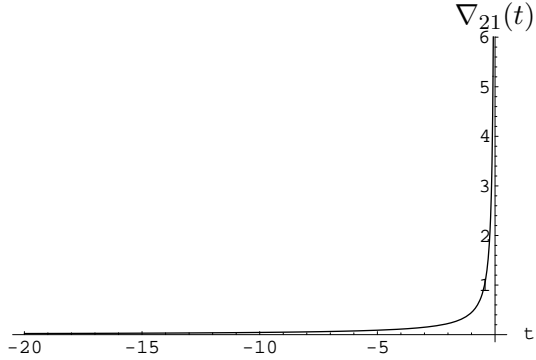}
\caption{The function $\nabla_{21}(t)$. Initial values at $\Lambda$ are 
$(a_{0},b_{0},c_{0})=(1000,1,1)$.}  
\label{sol.1}
\end{figure}

\begin{figure}
\psfrag{y}[][]{\small $(\hat{P_{22}}(t)-\hat{P_{11}}(t))$}
\includegraphics[scale=0.75]{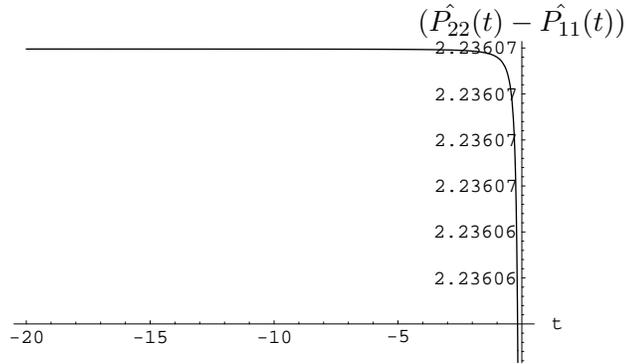}
\caption{The function $(\hat{P_{22}}(t)-\hat{P_{11}}(t))$. Initial values at
$\Lambda$ are $(a_{0},b_{0},c_{0})=(1000,1,1)$.}   
\label{sol.2}
\end{figure}

\begin{figure}
\psfrag{y}[][]{\small $Re(\hat{P_{21}}(t))$}
\includegraphics[scale=0.75]{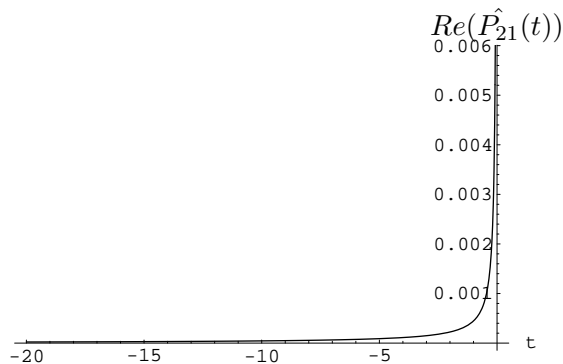}
\caption{The function $Re(\hat{P_{21}}(t))$. Initial values at $\Lambda$ are 
$(a_{0},b_{0},c_{0})=(1000,1,1)$}  
\label{sol.3}
\end{figure}

The evolutions of $\nabla_{ij}$, $(\hat{P_{ii}}-\hat{P_{jj}})$ and
$Re(\hat{P_{ij}})$ can be derived from Eq.~(\ref{rge:m}) and (\ref{rge:u}).
For $(i,j)=(2,1)$, with 
$\nabla_{21} \gg \nabla_{31} \simeq \nabla_{32} \simeq 1$, the RGE's for these
three functions form a {\it complete set} of coupled differential equations as
follows:   
\begin{subequations}
 \label{coupled}
 \begin{eqnarray}
  \frac{d\nabla_{21}}{dt} & = & \nabla_{21}^{2} (\hat{P_{22}}-\hat{P_{11}})
  \label{equationa}
 \\
  \frac{d(\hat{P_{22}}-\hat{P_{11}})}{dt} &
  = & -4\nabla_{21}[Re(\hat{P_{21}})]^{2} 
  \label{equationb}
 \\ 
  \frac{d Re( \hat{P_{21}})}{dt} & =& \nabla_{21} 
  ( \hat{P_{22}} - \hat{P_{11}})  Re( \hat{P_{21}} ) 
  \label{equationc}
 \end{eqnarray}
\end{subequations}
The exact solutions to these coupled differential equations are given by
\begin{subequations}
 \label{sol}
 \begin{eqnarray}
  \nabla_{21}(t) & = & a_{0}Z(t)^{-1/2}
  \label{equationa}
 \\
  (\hat{P_{22}}(t)-\hat{P_{11}}(t)) & =
   & (b_{0}^{2}+4c_{0}^{2}(1-Z(t)^{-1}))^{1/2} 
  \label{equationb}
 \\
  Re(\hat{P_{21}}(t)) & = & c_{0}Z(t)^{-1/2}
  \label{equationc}
 \end{eqnarray}
\end{subequations}
where
\begin{equation}
Z(t)\equiv 1-2a_{0}b_{0}t+a_{0}^{2}(b_{0}^{2}+4c_{0}^{2})t^{2}
\end{equation}
and $a_{0}$, $b_{0}$ and $c_{0}$ are the initial 
values at the high energy scale $\Lambda$:
\begin{eqnarray}
a_{0} & \equiv & \nabla_{21}(0); \quad
b_{0}  \equiv (\hat{P_{22}}(0)-\hat{P_{11}}(0))\nonumber\\
c_{0} & \equiv & Re(\hat{P_{21}}(0)).
\end{eqnarray}
The behaviours of these three functions are shown in 
Fig.~(\ref{sol.1})-(\ref{sol.3}).
Note that $\nabla_{21}(t)$ and $Re(\hat{P_{21}}(t))$ flow to zero, while
$(\hat{P_{22}}(t)-\hat{P_{11}}(t))$ flows to a constant value of
$(b_{0}^2 + 4c_{0}^2)^{1/2}$ in the infrared. This set of parameters,
$(\nabla_{21}(t^{\ast}), Re(\hat{P_{21}}(t^{\ast})),
(\hat{P_{22}}(t^{\ast})-\hat{P_{11}}(t^{\ast}))) 
= (0,0,(b_{0}^2 + 4c_{0}^2)^{1/2})$ is an infrared stable fixed point;
however, it is unrealistic. The function $\nabla_{21}(t)$ decreases to $O(1)$
very fast as the energy scale goes down, for any non-vanishing $b_{0}$ and
$c_{0}$, however small they are. This is phenomenologically unacceptable. In
addition, it contradicts with the assumption $\nabla_{21} \gg \nabla_{32},
\nabla_{31}$ we made in order to arrive at Eq.~(\ref{coupled}). For the
consistency of the calculations, we thus require the following two conditions
at the initial high energy scale $\Lambda$:   
\begin{eqnarray} b_{0}  &\equiv&
(\hat{P_{22}}(0)-\hat{P_{11}}(0)) = 0 \label{relation1}\\ 
c_{0} & \equiv &
Re(\hat{P_{21}}(0)) = 0 
\label{relation2} 
\end{eqnarray}
We emphasize that these conditions have been obtained by demanding that the
exact solutions to the above RGE's Eq.~(\ref{coupled}) be consistent with 
$\nabla_{21} \gg 1$. It is to be noted that these conditions have been 
obtained before numerically and perturbatively 
\cite{Casas:1999tg,Casas:1999ac,Chankowski:1999xc}.  
When these conditions are satisfied, all three equations
in Eq.~(\ref{coupled}) do not evolve. The first relation,
Eq.~(\ref{relation1}), gives rise to $|V_{32}|^{2} = |V_{31}|^{2}$ which
translates into \begin{equation} \frac{c_{1}^{2}s_{2}^{2}-s_{1}^{2}}{\sin
2\theta_{1} \cdot s_{2}} = \tan 2\theta_{3} \cdot \cos \delta 
\end{equation}
The second relation, Eq.~(\ref{relation2}), gives rise to
\begin{eqnarray}
\frac{Re(V_{32}^{\ast}V_{31})}{Im(V_{32}^{\ast}V_{31})}
& = & \tan(\frac{\phi^{'}-\phi}{2})\\
& = & \frac{\sin 2\theta_{3}
(c_{1}^{2} s_{2}^{2} - s_{1}^{2})}
{\sin \delta \cdot \sin 2\theta_{1} \cdot s_{2}} 
+ \frac{\cos 2\theta_{3}}{\tan \delta} \nonumber 
\end{eqnarray}
Combining these two relations, we obtain a very simple relation among
$\theta_{3}$ and three CP violating phases $\delta, \phi, \phi^{'}$:    
\begin{equation}
\cos 2\theta_{3} = -\frac{1}{\tan \delta} \cdot 
\frac{1}{\tan(\frac{\phi-\phi^{'}}{2})}
\end{equation}
We have studied the RGE's involving various functions, 
$\nabla_{ij}$, $(\hat{P_{ii}}-\hat{P_{jj}})$ and $Re(\hat{P_{ij}})$,
for the case $(i,j)=(3,1)$ and $(3,2)$.
Upon imposing the above consistency conditions Eq.~(\ref{relation1}) and
(\ref{relation2}), we deduce that the functions
$\hat{P_{11}}$, $\hat{P_{22}}$, $\hat{P_{33}}$, $Re(\hat{P_{31}})$ and
$Re(\hat{P_{32}})$ do not run. These results cannot be tested experimentally
at present. 

Now we discuss the implications of Eq.~(\ref{relation1}) and (\ref{relation2}) 
in the limit $\theta_{2} = 0$ (Recently, it has been pointed out that this
could be a consequence of the so-called 2-3 symmetry \cite{Lam:2001fb}). 
They imply
\begin{equation} \label{condition}
\cos (\frac{\phi-\phi^{'}}{2}) s_{1}^2 c_{3} s_{3} = 0, \qquad
s_{1}^{2} (c_{3}^{2} - s_{3}^{2}) = 0
\end{equation}
Since the atmospheric angle $\theta_{1} = \pi / 4$ is non-vanishing, 
these two relations can be satisfied simultaneously only if (i)
the solar mixing angle is maximal, i.e. $\theta_{3} = \pi / 4$, and (ii)
the Majorana phase difference $(\phi-\phi^{'})=\pi$.
The phases $\phi$ and $\phi^{'}$ occur in the
matrix element $\left< M_{ee} \right>$  for the neutrinoless double beta decay:
\begin{eqnarray}
\left< M_{ee} \right> &\equiv&| \sum_{i=1,2,3} U_{ei}^{2} m_{i} |
\\
&=&|m_{1} e^{-i\phi} c_{2}^{2} c_{3}^{2} + 
m_{2} e^{-i\phi^{'}} c_{2}^{2} s_{3}^{2} + 
m_{3} s_{2}^{2} e^{-2i\delta} | < B \nonumber
\end{eqnarray}
where index $i$ denotes the mass eigenstates.
Currently, the most stringent bound is given by $B=0.2 eV$
\cite{Baudis:1999xd}. In the limit $\theta_{2}=0$ with nearly degenerate
$m_{1} \simeq m_{2}$, $M_{ee}$ becomes
\begin{equation}
\label{0nub}
M_{ee} \simeq | m_{1} c_{2}^{2} ( e^{-i\phi} c_{3}^{2} +
e^{-i\phi^{'}} s_{3}^{2}) |
\end{equation}
It is obvious that when $(\phi-\phi^{'})=\pi$ and $\theta_{3}=\pi/4$ 
the r.h.s. of Eq.~(\ref{0nub}) is exactly zero. Thus we conclude that 
neutrinoless double beta decay is very highly suppressed.

It is interesting to speculate the reason(s) for consistency conditions of
Eq.~(\ref{relation1}) and (\ref{relation2}). It could be due to the existence
of a symmetry at a high energy scale $\Lambda$. 
The other possibility is that these two relations are the fixed point
relations of the RGE's for new physics above the scale $\Lambda$. 

% figures should be put into the text as floats.
% Use the graphics or graphicx packages (distributed with LaTeX2e).
% See the LaTeX Graphics Companion by Michel Goosens, Sebastian Rahtz,
% \begin{figure}
% \caption{}
% \label{}
% \end{figure}


% tables follow here or maybe be put in the text
% The ruledtabular enviroment adds doubled rules to table and sets a
% nice set of default table settings.
% Use the table* environment to get a full-width table in two-column
% \begin{table}
% \caption{}
% \label{}
% \begin{ruledtabular}
% \end{tabular}
% \end{ruledtabular}
% \end{table}

\begin{acknowledgments}
This work was supported, in part, by the U.S. Department of Energy under grant
number  DE FG03-05ER40894.
\end{acknowledgments}

\bibliography{ref2}

\end{document}